\documentstyle[12pt]{article}

\begin{document}

\title{Initial Conditions for Smooth Hybrid Inflation}
\author{{\bf G. Lazarides, C. Panagiotakopoulos} \\
Physics Division\\
School of Technology\\
University of Thessaloniki\\
Thessaloniki 540 06, Greece\\
{\bf N.D.Vlachos}\\
Dept. of Theoretical Physics\\
University of Thessaloniki\\
Thessaloniki 540 06, Greece}
\date{}
\maketitle

\begin{abstract}
We perform a numerical investigation of the field evolution in
the smooth hybrid inflationary model. We find that for almost all
the examined initial values we do get an adequate amount of 
inflation. Our results show that the model is ''natural'' 
and satisfactory.
\end{abstract}

\baselineskip = .30in \vspace{.2in}

\newpage

The hybrid inflationary scenario$^{(1)}$ proposed by Linde in the 
context of non supersymmetric (SUSY) theories is a realization 
of chaotic inflation based on a coupled system of two scalar fields
one of which may not be a gauge singlet. The great advantage of 
this scenario is that it produces the observed temperature
fluctuations in the cosmic background radiation (CBR)
with natural values of the coupling constants. However, inflation
terminates abruptly and is followed by a ''waterfall'' regime during
which topological defects can be easily produced. Recently, two of
us have proposed$^{(2)}$ a variant of Linde's potential which can
be derived in a wide class of SUSY grand unified theories (GUTs)
based on semi-simple gauge groups by utilizing the first
non-renormalizable contribution to the superpotential. Although
one gets only a slight variation of Linde's potential, the
cosmological scenario obtained is drastically different. Already
since the beginning of inflation, the system follows a particular
valley of minima which leads to a particular point of the vacuum
manifold. Thus, this inflationary scenario does not lead to
production of topological defects. Also the termination of inflation
is not as abrupt as in the hybrid case. It is quite smooth and
resembles more the cases of new or chaotic inflation. The main
advantage of this smooth hybrid inflationary scenario is that the
measured value of the temperature fluctuations of CBR can be
reproduced with natural values of the parameters and with a GUT
scale, $M_X$, consistent with the unification of the minimal
supersymmetric standard model (MSSM) gauge couplings. It is also
remarkable that the scale controlling the non-renormalizable terms
in the superpotential turns out to be of order 10$^{18}$ GeV.
The spectral index of density fluctuations is close to unity.

For an inflationary scenario to be considered fully successful,
one has to show that it is obtainable for a wide class of ''natural''
initial values of the fields and their time derivatives. In ref.(2),
two sets of initial conditions were studied semianalytically and it
was argued that they lead to smooth hybrid inflation. These sets,
however, cannot be considered completely ''natural'' since they
require some or even a considerable discrepancy between the initial
values of the fields. The purpose of this paper is to identify a wide
class of ''natural'' initial conditions for smooth hybrid inflation,
i.e., comparable initial values of the fields for which the system
falls at the bottom of a particular valley of minima. Its
subsequent evolution along this valley, then, produces smooth hybrid
inflation. To this end, we solve numerically the evolution equations
of the system for a wide class of initial conditions. The result is
striking and unexpected. We find that, for almost all the examined
initial conditions (except a narrow transition region), we do get
smooth hybrid inflation with an adequate number of e-foldings.
This result together with the other advantages of this inflationary
scenario makes it fully satisfying and ''natural''. Our analysis
also includes cases with all initial field values being much smaller
than the Planck scale. In these cases, our results are expected to
be less affected by replacing global by local supersymmetry.

The smooth hybrid inflationary scenario can be realized in the
context of a SUSY GUT based on a gauge group $G$ of rank $\geq 5$.
We assume that $G$ breaks spontaneously directly to the standard
model (SM) group $G_S$ at a scale $M_X\sim 10^{16}$ GeV and that
below $M_X$ the only SM non-singlet states of the theory are the
usual MSSM states. This guarantees the successful MSSM predictions
for $sin^2\theta _w$ and $\alpha _s$. The theory could also possess
some global symmetries. The symmetry breaking of $G$ to $G_S$
is obtained through a superpotential which includes the terms 
\begin{equation}
W=s(-\mu ^2+\frac{(\phi \bar \phi )^2}{M^2}).
\end{equation}
Here $\phi \,$,$\bar \phi $ is a conjugate pair of left-handed SM
singlet superfields which belong to non-trivial representations
of the gauge group $G$ and reduce its rank by their vacuum
expectation values (vevs), $s$ is a gauge singlet left-handed
superfield, $\mu $ is a superheavy mass scale related to $M_X$,
whereas $M$ is a mass scale of the order of the ''compactification''
scale $M_c\sim 10^{18}$ GeV which controls the
non-renormalizable terms in the superpotential of the theory. The
superpotential terms in eq.(1) are the dominant couplings involving
the superfields $s\,$,$\phi \,$,$\bar \phi $ consistent with a
continuous R-symmetry under which
$W\rightarrow e^{i\theta }W$,$s\rightarrow e^{i\theta}s$,
$\phi \bar \phi \rightarrow \phi \bar \phi $ and a discrete symmetry
under which $\phi \bar \phi $ changes sign.

The potential obtained from $W$ in eq.(1), in the supersymmetric
limit, is 
\begin{equation}
V=|\mu ^2-\frac{(\phi \bar \phi )^2}{M^2}|^2+4|s|^2
\frac{|\phi |^2|\bar \phi
|^2}{M^4}(|\phi |^2+|\bar \phi |^2)+\textstyle{D-terms},
\end{equation}
where the scalar components of the superfields are denoted by the
same symbols as the corresponding superfields. Vanishing of
the D-terms is achieved along the D-flat directions where
$|\bar \phi |=|\phi |$. The supersymmetric vacuum 
\begin{equation}
<s>=0,\>\><\phi ><\bar \phi >=\pm \mu M,\>\>|<\bar \phi >|=|<\phi >|
\end{equation}
lies on the particular D-flat direction $\bar \phi ^{*}=\pm \phi $.
Restricting ourselves to this direction and performing appropriate
gauge,discrete and R-transformations we can bring the complex
$s$,$\phi $,$\bar\phi $ fields on the real axis, i.e.,
$s\equiv \frac \sigma {\sqrt{2}}$,
$\>\bar \phi =\phi \equiv \frac 12\chi $, where $\sigma $
and $\chi $ are real scalar fields. The potential in eq.(2) then
takes the form 
\begin{equation}
V(\chi ,\sigma )=(\mu ^2-\frac{\chi ^4}{16M^2})^2+\frac{\chi ^6
\sigma ^2}{16M^4}
\end{equation}
and the supersymmetric minima correspond to 
\begin{equation}
|<\chi >|=2(\mu M)^{1/2},\>\><\sigma >=0\,\ .
\end{equation}
The mass acquired by the gauge bosons is $M_X=g(\mu M)^{1/2}$,
where $g$ is the GUT gauge coupling. For any fixed value of
$\sigma $, the potential in
eq.(4), as a function of $\chi ^2$, has a local maximum at
$\chi ^2=0$ and an absolute minimum lying at 
\begin{equation}
\chi ^2\simeq \frac 43\frac{\mu ^2M^2}{\sigma ^2}
\ ,\>\textstyle{for}\>\sigma^2>>\mu M\ ,
\end{equation}
and $\chi ^2\simeq 4\mu M$, for $\sigma ^2<<\mu M$. The value of the
potential along the maxima at $\chi ^2=0$ is constant, $V_{max}(\chi
^2=0)=\mu ^4$.

Assume for the moment that at some region of the universe the scalar
fields $\chi $ and $\sigma $, starting from appropriate initial
conditions to be discussed below, evolve in such a way so they
become almost uniform with values at the bottom of the valley of
minima in eq.(6). One can then show$^{(2)}$ that these fields,
at subsequent times, move towards the supersymmetric minima in
eq.(5) following this valley and the system inflates till $\sigma $
reaches the value 
\begin{equation}
\sigma \simeq \sigma _o\equiv (\frac{2M_P}{9\sqrt{\pi }
(\mu M)^{1/2}})^{1/3}(\mu M)^{1/2}\stackrel{_{>}}{_{\sim }}
(\mu M)^{1/2},
\end{equation}
where $M_P=1.2\times 10^{19}$ GeV is the Planck mass. The number of
e-foldings from the moment at which the $\sigma $ field has the
value $\sigma $ till the end of inflation is given by 
\begin{equation}
N(\sigma )\simeq (\frac{3\sqrt{2\pi }}{2\mu MM_P})^2\sigma ^6.
\end{equation}
This implies that the value of the $\sigma $ field when the present
horizon crossed outside the inflationary horizon was 
\begin{equation}
\sigma _H\simeq (\frac{9N_H}2)^{1/6}\sigma _o\,,
\end{equation}
where $N_H$ is the number of e-foldings of the present horizon size
during inflation. After the end of inflation, the $\sigma $
and $\chi $ fields enter smoothly into an oscillatory phase about
the global supersymmetric minimum of the potential in eq.(5) with
frequency $m_\sigma =m_\chi =2\sqrt{2}(\mu /M)^{1/2}\mu $. These
fields should eventually decay into lighter particles and ''reheat''
the universe. Taking $N_H=60$, $M_X=2\times 10^{16}$
GeV, $g=0.7$ (consistent with the MSSM unification),the microwave
background quadrupole anisotropy $(\Delta T/T)\simeq 5\times 10^{-6}$
from the Cosmic Background Explorer (COBE) implies that
$M\simeq 9.4\times 10^{17}$ GeV and $\mu \simeq 8.7\times 10^{14}$
GeV. With these numbers, which we will use throughout this paper,
one estimates the value of $\sigma $ at the end of inflation
(see eq.(7)) $\sigma _o\simeq 1.1\times 10^{17}$ GeV $
\simeq 5.4M_X$ and its value when the present horizon size
crossed outside the inflationary horizon (see eq.(9))
$\sigma _H\simeq 2.54\sigma _o\simeq 2.7\times 10^{17}$ GeV.
An important advantage of the above smooth hybrid inflationary
scenario is that inflation takes place at relatively low values
of the field $\sigma $ and, therefore, there is hope that it
survives even in the context of supergravity theories although
it may acquire drastic modifications.

We will now try to specify the initial conditions for the $\sigma $
and $\chi $ fields which lead to the above described inflationary
scenario. In other words, we will try to identify the initial
conditions for which the system falls at the bottom of the valley
of minima in eq.(6) with a value of $\sigma \geq \sigma _H$ so that
its subsequent evolution along the valley
produces an adequate amount of inflation. We assume that, after
''compactification'' at some initial cosmic time, a region emerges
in the universe where the scalar fields $\sigma $ and $\chi $
happen to be almost uniform with negligible kinetic energies.
(The initial values of $\sigma $ and $\chi $ can always be
transformed by appropriate gauge and R-transformations to become
positive.) The evolution of the system in this
region is governed by the following equations of motion 
\begin{equation}
\ddot \chi +3H\dot \chi -\frac{\chi ^3}{2M^2}(\mu ^2-
\frac{\chi ^4}{16M^2})+\frac{3\chi ^5\sigma ^2}{8M^4}=0\ ,
\end{equation}
\begin{equation}
\ddot \sigma +3H\dot \sigma +\frac{\chi ^6\sigma }{8M^4}=0\ ,
\end{equation}
where overdots denote derivatives with respect to cosmic time
and $H$ is the Hubble parameter 
\begin{equation}
H=(\frac{8\pi }3)^{1/2}M_P^{-1}\varrho ^{1/2}=(\frac{8\pi }%
3)^{1/2}M_P^{-1}(\frac 12\dot \chi ^2+\frac 12\dot
\sigma ^2+V(\chi ,\sigma))^{1/2}
\end{equation}
($\varrho $ is the energy density).

The case where initially $\sigma \gg M_P\gg \chi $ was examined
in ref.(2).Under these circumstances, the last term in eq.(4) is
initially the dominant contribution to the potential energy density
of the system (assuming $\chi \stackrel{_{>}}{_{\sim }}
(\mu ^2M^2/\sigma )^{1/3}$). Also, eq.(10) reduces to 
\begin{equation}
\ddot \chi +3H\dot \chi +\frac{3\chi ^5\sigma ^2}{8M^4}\simeq 0.
\end{equation}
Let us for the moment assume that $\sigma $ remains almost
constant. The frequency of oscillations of the $\chi $ field is
then much greater that $H$ and $\chi $ initially performs damped
oscillations over the maximum at $\chi=0$. The continuity equation 
\begin{equation}
\dot \varrho =-3H(\varrho +p),
\end{equation}
with $p$ being the pressure averaged over one oscillation of
$\chi $ becomes 
\begin{equation}
\dot \varrho =-3H\gamma \varrho ,
\end{equation}
where $\gamma =3/2$ for a $\chi ^6$ potential$^{(3)}$. This equation
together with the fact that $\varrho $ is proportional to $H^2$
gives
\begin{equation}
H\simeq \frac 4{9t}.
\end{equation}
Comparing this result with eq.(12) we can obtain the amplitude of
the oscillating $\chi $ field 
\begin{equation}
\chi _m\simeq (\frac 3{8\pi })^{1/6}(\frac{16M^2M_P}{9\sigma t})^{1/3}.
\end{equation}
Eq.(11) averaged over one oscillation of $\chi $ then gives 
\begin{equation}
\ddot \sigma +\frac 4{3t}\dot \sigma +\frac 1{27\pi t^2}(\frac{M_P}
\sigma)^2\sigma \simeq 0.
\end{equation}
The solutions of this equation are of the form $\sigma =t^\alpha $,
 where $\alpha $ satisfies the quadratic equation 
\begin{equation}
\alpha ^2+\frac 13\alpha +\frac 1{27\pi }(\frac{M_P}\sigma )^2\simeq 0.
\end{equation}
For $\sigma \gg M_P$, the solutions are 
\begin{equation}
\alpha \simeq -\frac 13+\frac 1{9\pi }(\frac{M_P}\sigma )^2
\quad \mbox{and} \>\>\alpha \simeq -
\displaystyle\frac 1{9\pi }(\frac{M_P}
\sigma )^2.
\end{equation}
This means that $\sigma $ quickly approaches an extremely slowly
decreasing function of time and, thus, our starting assumption
that it remains approximately constant is justified. When the
amplitude of the $\chi $ field drops to about $(\mu ^2M^2/
\sigma )^{1/3}$, the $\mu ^4$ term dominates the
potential in eq.(4) and the Hubble parameter becomes approximately
constant and equal to $H=(8\pi /3)^{1/2}\mu ^2/M_P$ and remains so
thereafter till the end of inflation. The subsequent evolution of
the system has been studied in detail in ref.(2). The overall
conclusion is that,in a time interval 
\begin{equation}
\Delta t\sim 6\pi (\frac \sigma {M_P})^2H^{-1},
\end{equation}
the $\chi $ field falls into the valley of minima in eq.(6) and
relaxes at the bottom of this valley whereas the $\sigma $ field
still remains unchanged and much greater than $M_P$. After that,
the system follows the valley of minima towards the supersymmetric
vacuum and, therefore,the smooth hybrid inflationary scenario is
realized for initial values of the fields satisfying the inequality
$\sigma \gg M_P\gg \chi $. However,these initial conditions cannot
be considered totally satisfying because, assuming that the initial
energy density is well below $M_P^4$ , we see that there must be
some discrepancy between the initial values of the fields. Also the
inclusion of supergravity is expected to invalidate the above
discussion of initial conditions which involve values of the field
$\sigma \stackrel{_{>}}{_{\sim }}M_P$. To minimize the influence
from supergravity one could start with field values much smaller
than $M_P$ and, as pointed out in ref.(2),still obtain adequate
inflation. However, this required an initial energy density much
smaller than $M_P^4$ and an initial value of $\chi $
''unnaturally'' smaller than the initial value of $\sigma $.

For smooth hybrid inflation to be considered as a fully successful
inflationary scenario, one must show that it is obtained for a wide
class of initial conditions which are more ''natural'' than the
ones just discussed.This can be done only numerically. To this end,
we put $\hat \chi \equiv \chi /M_P$ and $\hat \sigma \equiv
\sigma /M_P$ in eqs. (10) and (11) which become 
\begin{equation}
\hat \chi ^{\prime \prime }+3\hat H\hat \chi ^{\prime }-
\frac{\hat \chi ^3}{32\hat M^4}(16\hat \mu ^2\hat M^2-
\hat \chi ^4)+\frac{3\hat \chi ^5\hat \sigma ^2}{8\hat M^4}=0\ ,
\end{equation}
\begin{equation}
\hat \sigma ^{\prime \prime }+3\hat H\hat \sigma ^{\prime }+
\frac{\hat \chi^6\hat \sigma }{8\hat M^4}=0\ ,
\end{equation}
where 
\begin{equation}
\hat H\equiv \frac H{M_P}=(\frac{8\pi }3)^{1/2}[\frac 12
(\hat \chi ^{\prime})^2+\frac 12(\hat \sigma ^{\prime })^2+
\hat V(\hat \chi ,\hat \sigma)]^{1/2}
\end{equation}
with 
\begin{equation}
\hat V(\hat \chi ,\hat \sigma )\equiv \frac{V(\chi ,\sigma )}
{M_P^4}=(\hat\mu ^2-\frac{\hat \chi ^4}{16\hat M^2})^2+
\frac{\hat \chi ^6\hat \sigma ^2}{16\hat M^4}\ \cdot
\end{equation}
Here primes denote derivatives with respect to the dimensionless time
variable $\tau \equiv M_P\,t$ and 
\begin{equation}
\hat M\equiv \frac M{M_P}\simeq \frac 1{12.83},\>\>\hat \mu
\equiv \frac \mu{M_P}\simeq \frac 1{13737}.
\end{equation}
We have conducted numerical integration of eqs.(22) and (23) for an
extensive set of initial values of the fields in the ranges
$0.01\leq \hat \sigma \leq 1.2$ and $0.01\leq \hat \chi \leq 0.5$
and with vanishing initial velocities. The integration of these two
coupled equations was performed by implementing a variant of the
Bulrish-Stoer$^{(4)}$ variable step method in a Fortran program run
mainly on a number of workstations. As a general rule, the initial
step for the dimensionless time variable $\tau $ was chosen to be 10
while the sought accuracy was put to $10^{-12}$. This choice was
found to ensure reasonable stability in the cases where $\hat\sigma $
settles down to a constant non-zero value relatively early. In the
opposite cases as well as near some transition regions, the initial
step was decreased to 5 or less, while the sought accuracy was
increased by two or three orders of magnitude depending on the case.

The results of our search are summarized in figures 1 and 2 .
Each point shown on the $\widehat{\sigma }$-$\widehat{\chi }$
plane corresponds to a given set of initial conditions and depicts
a definite evolution pattern for the $\widehat{\sigma }$-
$\widehat{\chi }$ system according to the symbol attributes being
used. We have used filled circles to specify the evolution pattern
where both fields oscillate and fall rapidly to the supersymmetric
minima in eq.(5) without producing any appreciable amount of
inflation. Open triangles correspond to the case where
$\widehat{\sigma }$ starts-off at relatively large values
($\widehat{\sigma }>\widehat{\chi }$) and decreases
slowly tending asymptotically to a constant value
$\widehat{\sigma }\geq \widehat{\sigma }_H\approx 0.0225$ .
The field $\widehat{\chi }$ oscillates and relaxes at the bottom
of the valley in eq.(6). The system then evolves through the valley
of minima in eq.(6) giving an adequate amount of inflation. Finally,
open circles correspond to the pattern where both fields
start oscillating at the beginning. Then, $\widehat{\sigma }$
settles down at large values and the system subsequently follows
the valley of minima in eq.(6) as in the preceding case.

The evolution pattern represented by open triangles includes the
pattern found in the limiting case $\widehat{\sigma }\gg 1\gg
\widehat{\chi }$ analyzed earlier by means of semianalytic
arguments. In general, however, $\widehat{\sigma }$ does not remain
frozen as in the limiting case, but its variation over a large
period of time is small. The open circles area being the least
expected deserves further attention. Here, although
$\widehat{\sigma }$ starts at moderate or small values, it appears
to increase in amplitude absorbing energy from the fast oscillating
field $\widehat{\chi }$ and creates conditions that eventually lead
to evolution of the open triangles type. It is a beautiful example
of large energy transfer between two strongly-coupled non-linear
oscillators.

The points depicted in figs.1 and 2 follow a remarkably regular
pattern, although there is some intermingling. This intermingling
persisted even when the sought accuracy was increased to a maximum
and the step was lowered to a minimum, thus, we have to assume that
it really exists. It would, of course,be very desirable to get
three distinct regions separated by critical lines but we have no
reason to a priori exclude the possibility of intermingling.

The problem to be addressed next concerns the ''naturality''
of initial conditions. A closer look at fig.1 reveals that the open
triangles area,although leading to successful inflation, cannot
be considered as being completely ''natural'' since it requires
relatively large differences between the initial values of the
fields. The open circles region appears to be significantly better
since the initial values can be of the same order of magnitude.
Considering that the initial kinetic energies for both fields are
taken to be zero, the initial energy density equals the potential
energy density given in eq.(25). It turns out that, when
$\widehat{\textstyle{ }\chi }$ and $\widehat{\sigma }$ are of the
same order of magnitude and approximately equal to a few tenths,
we get acceptable initial energy densities. Thus,part of the open
circles area corresponds to ''natural'' initial conditions
which lead to successful inflation. Inclusion of supergravity will
certainly invalidate the preceding analysis, except for an area
where all initial field values are much smaller that the Planck
scale. This area is shown in fig.2 and a closer inspection shows
that all but one of the points fulfil all the conditions for
adequate inflation. In summary, taking into account
all the previously stated results, we feel confident to conclude
that the smooth hybrid inflationary model appears to be ''natural''
and satisfactory.

{\it Acknowledgments. }We would like to express our gratitude to our
colleagues of the Astrophysics Department and especially to Prof. K.
Kokkotas, for sharing their computer facilities with us. This work is
supported in part by the E.U. Science project SC1-CT91-0729.

\section*{References}

\begin{enumerate}
\item  A. D. Linde, Phys. Rev. \underline{D49} (1994) 748.

\item  G. Lazarides and C. Panagiotakopoulos, Phys. Rev.
\underline{D52} (1995) R559

\item  M. Turner, Phys. Rev. \underline{D28} (1983) 1243.

\item  Numerical Recipes in Fortran: The Art of Scientific Computing
(Cambridge University Press, Second Edition 1992).
\end{enumerate}

\newpage

\section*{Figure Captions}

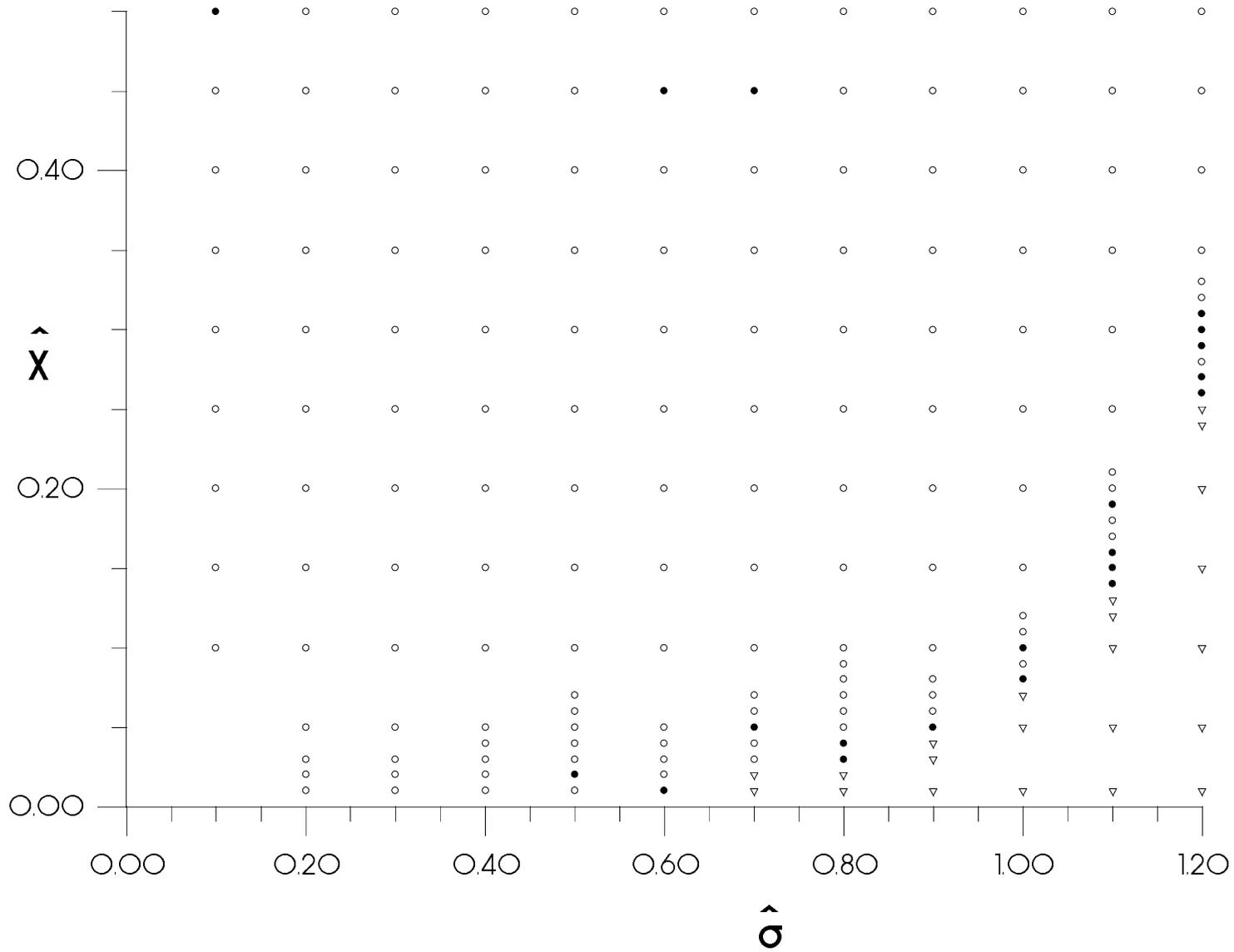
{\footnotesize \noindent {\bf Fig.~1}: Evolution patterns for the }
$\widehat{\sigma }\textstyle{-}\widehat{\chi }$ {\footnotesize system.
Filled circles represent points that do not lead to inflation.
Open triangles give adequate inflation having only the }
$\widehat{\chi }$ {\footnotesize field
oscillating. Open circles give adequate inflation with both fields
oscillating initially. }

{\footnotesize \vspace{1cm} }

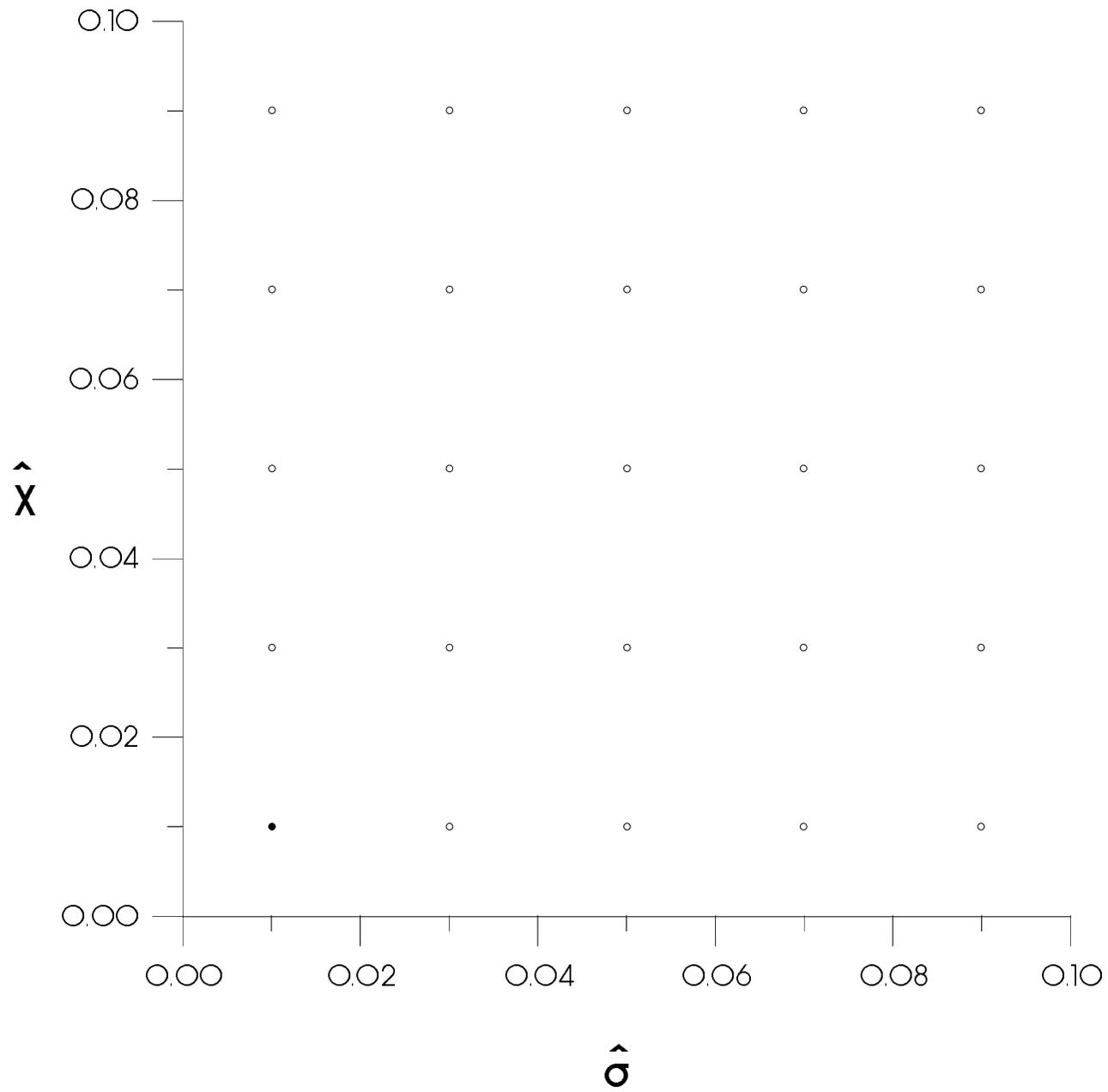
{\footnotesize \noindent {\bf Fig.~2}: Same as in Fig.1. The initial values
are now restricted to lie near the beginning of the axes. }

\end{document}